\documentclass[12pt]{article}
\usepackage{amsmath,amssymb,amsthm,amsxtra,overpic,bbm,bm,epsfig,subfigure}
\usepackage{color}
\usepackage{comment}

\usepackage{slashed,stmaryrd}

\begin{document}

\vspace{0.2cm}

\begin{center}
{\large\bf Discriminating between Thermal and Nonthermal Cosmic Relic Neutrinos through Annual Modulation at PTOLEMY}
\end{center}

\vspace{0.2cm}

\begin{center}
{\bf Guo-yuan Huang $^{a, b}$}, \footnote{E-mail: huanggy@ihep.ac.cn}
\quad
{\bf Shun Zhou $^{a, b, c}$} \footnote{E-mail: zhoush@ihep.ac.cn}
\\
{\small $^a$Institute of High Energy Physics, Chinese Academy of
Sciences, Beijing 100049, China \\
$^b$School of Physical Sciences, University of Chinese Academy of Sciences, Beijing 100049, China \\
$^c$Center for High Energy Physics, Peking University, Beijing 100871, China}
\end{center}

\vspace{1.5cm}

\begin{abstract}
If massive neutrinos are Dirac particles, the proposed PTOLEMY experiment will hopefully be able to discover cosmic neutrino background via $\nu^{}_e + {^3}{\rm H} \to {^3}{\rm He} + e^-$ with a capture rate of $\Gamma^{}_{\rm D} \approx 4~{\rm yr}^{-1}$. Recently, it has been pointed out that right-handed components of Dirac neutrinos could also be copiously produced in the early Universe and become an extra thermal or nonthermal ingredient of cosmic relic neutrinos, enhancing the capture rate to $\Gamma^{}_{\rm D} \approx 5.1~{\rm yr}^{-1}$ or $\Gamma^{}_{\rm D} \approx 6.1~{\rm yr}^{-1}$. In this work, we investigate the possibility to distinguish between thermal and nonthermal spectra of cosmic relic neutrinos by measuring the annual modulation of the capture rate. For neutrino masses of $0.1~{\rm eV}$, we have found the amplitude of annual modulation in the standard case is ${\cal M} \approx 0.05\%$, which will be increased to $0.1\%$ and $0.15\%$ in the presence of additional thermal and nonthermal right-handed neutrinos, respectively. The future detection of such a modulation will be helpful in understanding the Majorana or Dirac nature of massive neutrinos.
\end{abstract}

\begin{flushleft}
\hspace{0.8cm} PACS number(s): 98.62.Sb, 14.60.St, 25.30.Pt
\end{flushleft}

\def\thefootnote{\arabic{footnote}}
\setcounter{footnote}{0}

\newpage

\section{Introduction}

The discovery and a subsequent precision measurement of cosmic microwave background (CMB) have provided us with precious information on the evolution of our Universe, and helped us to establish the standard model of cosmology~\cite{Agashe:2014kda}. In contrast, only an indirect evidence for the existence of cosmic neutrino background (C$\nu$B) has been found in the observations of CMB, primordial abundances of light nuclear elements and cosmic large-scale structures, where neutrinos have played a crucially important role~\cite{Kolb,Dodelson,XZ,Lesgourgues}. The direct detection of C$\nu$B in the future will be another great triumph of the standard cosmology, and also offer a new possibility to probe the intrinsic properties of massive neutrinos.

According to the standard cosmology, the temperature of C$\nu$B at present is extremely low, namely, $T^{}_\nu \approx 1.95~{\rm K}$, and the corresponding average momentum is $\langle p^{}_\nu \rangle \approx 5\times 10^{-4}~{\rm eV}$~\cite{Kolb,Dodelson,XZ,Lesgourgues}. Consequently, it will be an extremely difficult task to detect C$\nu$B, although the number density for each neutrino or antineutrino species can be as large as $\overline{n}^{}_\nu \approx 56~{\rm cm}^{-3}$. As was pointed out by Weinberg in 1962~\cite{Weinberg:1962zza}, cosmic relic neutrinos can be captured on beta-decaying nuclei, such as $\nu^{}_e + {^3}{\rm H} \to {^3}{\rm He} + e^-$, which can take place without an energy threshold for the initial-state neutrinos. In Ref.~\cite{Irvine:1983nr}, it was noticed that the signal of the capture of C$\nu$B will be a peak in the final-state electron spectrum, which is located at a distance of $2m^{}_\nu$ away from the end point of beta-decay spectrum of ${^3}{\rm H} \to {^3}{\rm He} + \overline{\nu}^{}_e + e^-$. Therefore, in order to observe this tiny separation, we need an energy resolution even smaller than the absolute neutrino mass $m^{}_\nu$, which is now known to be at or below the sub-eV level.

Recently, a lot of attention has been paid to the detection of C$\nu$B through the capture on beta-decaying nuclei. See, e.g.,  Refs.~\cite{Ringwald:2005zf} and \cite{Vogel:2015vfa}, for reviews on the other possible approaches. In Ref.~\cite{Cocco:2007za}, a detailed calculation of the capture rate for $\nu^{}_e + {^3}{\rm H} \to {^3}{\rm He} + e^-$ was first performed, and the impact of neutrino mass hierarchies and flavor mixing on the capture rate was investigated in Ref.~\cite{Blennow:2008fh}. In addition, similar processes for the eV- and keV-mass sterile neutrinos or for different nuclear isotopes have also been discussed in the literature~\cite{Liao:2010yx, Kaboth:2010kf, Li:2010sn, Li:2010vy, Lusignoli:2010eq, Li:2011ne, Li:2011mw, deVega:2011xh, Faessler:2011zz, Faessler:2016tjf}. More importantly, the PTOLEMY experiment has been proposed~\cite{Betts:2013uya}, in which a 100 gram of surface-deposition tritium source will be implemented and the energy resolution for the final-state electrons can reach $\Delta \approx 0.15~{\rm eV}$. Given this nominal experimental setup, Long {\it et al.} and Lisanti {\it et al.} have found that the capture rate is $\Gamma^{}_{\rm D} \approx 4~{\rm yr}^{-1}$ if massive neutrinos are Dirac particles, while
that for Majorana neutrinos is twice larger, i.e., $\Gamma^{}_{\rm M} \approx 8~{\rm yr}^{-1}$~\cite{Long:2014zva, Lisanti:2014pqa}. Although the capture rate is small, it is already interesting to have a realistic experiment that will be able to directly prove the existence of C$\nu$B. Moreover, the difference between the capture rates $\Gamma^{}_{\rm D}$ and $\Gamma^{}_{\rm M}$ will enable us to discriminate Dirac and Majorana nature of massive neutrinos. Such a discrimination is extremely important for exploring the origin of neutrino masses.

Motivated by the proposed PTOLEMY experiment, the authors of Refs.~\cite{Zhang:2015wua} and \cite{Chen:2015dka} have examined how the copious production and subsequent decoupling of right-handed neutrinos in the early Universe can affect the capture rate of cosmic relic neutrinos, assuming massive neutrinos to be Dirac particles. It has been shown that the capture rate can be increased at most to be $\Gamma^{}_{\rm D} \approx 5.1~{\rm yr}^{-1}$ or $\Gamma^{}_{\rm D} \approx 6.1~{\rm yr}^{-1}$, if right-handed Dirac neutrinos are produced thermally~\cite{Zhang:2015wua} or nonthermally~\cite{Chen:2015dka}, respectively. In this framework, an immediate question arises: is it possible to distinguish between thermal and nonthermal energy spectra of relic right-handed neutrinos? The main purpose of this short note is to answer this question, and show that a PTOLEMY-like experiment with a total statistics of $10^6$ neutrino events may do such a challenging job.

The remaining part of this work is structured as follows. In Section 2, we briefly review possible ways to generate right-handed neutrinos in the early Universe and the impact of thermal and nonthermal spectra of relic right-handed neutrinos on the detection of C$\nu$B. Following the general idea of Ref.~\cite{Safdi:2014rza}, where the gravitational focusing effects by the Sun are considered, we demonstrate in Section 3 that the annual modulation of the capture rate can be used to discriminate the velocity distributions of cosmic relic neutrinos. Finally, we give some concluding remarks in Section 4.

\section{Cosmic Relic Neutrinos}

If the standard model of elementary particles (SM) is extended with three right-handed neutrinos, which are singlets under the SM gauge symmetries, neutrinos acquire tiny Dirac masses in the same way as charged leptons and quarks do. More explicitly, the gauge-invariant Lagrangian can be written as
\begin{eqnarray}
{\cal L} = {\cal L}^{}_{\rm SM} + \sum_{\alpha}\overline{\nu^{}_{\alpha {\rm R}}} {\rm i} {\slashed \partial} \nu^{}_{\alpha {\rm R}} - \left[\sum_{\alpha,\beta} \left(Y^{}_\nu\right)^{}_{\alpha \beta} \overline{\ell^{}_{\alpha {\rm L}}} \tilde{H} \nu^{}_{\beta {\rm R}} + {\rm h.c.}\right] \; ,
\label{eq:lag}
\end{eqnarray}
with $\alpha$ and $\beta$ running over $e$, $\mu$ and $\tau$, where ${\cal L}^{}_{\rm SM}$ stands for the SM Lagrangian, $\ell^{}_{\alpha {\rm L}}$ and $\tilde{H} \equiv {\rm i}\sigma^{}_2 H^*$ are lepton and Higgs doublets, $Y^{}_\nu$ is the Dirac neutrino Yukawa coupling matrix.
After the spontaneous gauge symmetry breaking, the neutrino mass matrix is given by $M^{}_\nu = Y^{}_\nu \langle H \rangle$ with $\langle H \rangle \approx 174~{\rm GeV}$ being the vacuum expectation value of the Higgs field. In order to obtain sub-eV neutrino masses ${\cal O}(M^{}_\nu) \sim 0.1~{\rm eV}$, we have to require extremely small Yukawa couplings ${\cal O}(Y^{}_\nu) \sim 10^{-12}$, which exaggerates the hierarchy problem of fermion Yukawa couplings, when compared to the top-quark Yukawa coupling $y^{}_t \approx 1$. On the other hand, the lepton-number conservation or a global $U(1)$ symmetry should be imposed on the Lagarangian in Eq.~(\ref{eq:lag}) to forbid a Majorana mass term for right-handed neutrino singlets. In the following discussions, however, we focus on the cosmological implications of massive Dirac neutrinos. A satisfactory solution to these two theoretical problems is beyond the scope of this work.

Within the SM and standard cosmology, the production rate of right-handed neutrinos in the early Universe is always far below the Hubble expansion rate at any time~\cite{Antonelli:1981eg,Zhang:2015wua}. Therefore, it is quite reasonable to ignore right-handed neutrinos $\nu^{}_{\alpha {\rm R}}$ and left-handed antineutrinos $\overline{\nu}^{}_{\alpha {\rm L}}$ for $\alpha = e, \mu, \tau$ in the evolution of our Universe. Nowadays, we have only left-handed neutrinos $\nu^{}_{\alpha {\rm L}}$ and right-handed antineutrinos $\overline{\nu}^{}_{\alpha {\rm R}}$ as a cosmic background, for which the average number density is $\overline{n}^{}_\nu \approx 56~{\rm cm}^{-3}$. When new physics scenarios are considered, it is possible to produce both $\nu^{}_{\alpha {\rm R}}$ and $\overline{\nu}^{}_{\alpha {\rm L}}$ abundantly and reconcile them with cosmological observations, in particular the effective number of neutrino species $N^{}_{\rm eff}$ from the latest data of Big Bang Nucleosynthesis (BBN) and CMB. Generally speaking, there exist two typical scenarios.
\begin{itemize}
\item {\it Thermal Right-handed Dirac Neutrinos}---As is shown in Ref.~\cite{Zhang:2015wua}, if primordial magnetic fields are generated in the electroweak phase transition~\cite{Enqvist:1998fw}, even a rather small magnetic dipole moment of massive Dirac neutrinos $\mu^{}_\nu = 3\times 10^{-20} \mu^{}_{\rm B}$ (with $\mu^{}_{\rm B}$ being the Bohr magneton and $m^{}_\nu \sim 0.1~{\rm eV}$) leads to a thermal production of $\nu^{}_{\alpha {\rm R}}$ via significant  $\nu^{}_{\alpha {\rm L}} \to \nu^{}_{\alpha {\rm R}}$ conversions~\cite{Enqvist:1994mb}. Depending on the evolution of primordial magnetic fields, the decoupling temperature of $\nu^{}_{\alpha \rm R}$ can be well above the temperature of QCD phase transition $T^{}_{\rm QCD} \approx 200~{\rm MeV}$. Hence, a proper decoupling temperature $T^{}_{\rm R}$ can be chosen to evade the cosmological bound $N^{}_{\rm eff} = 3.14^{+0.44}_{-0.43}$ at the $95\%$ confidence level~\cite{Ade:2015xua}. In this scenario, $\nu^{}_{\alpha {\rm R}}$ and $\overline{\nu}^{}_{\alpha {\rm L}}$ will survive today as extra thermal relics, whose distribution functions in the C$\nu$B frame are isotropic and can be parametrized as follows
    \begin{eqnarray}
    f^{}_{\rm TH}(p^{}_\nu) = \frac{1}{\exp(p^{}_\nu/T^0_{\rm R}) + 1} \; ,
    \label{eq:th}
    \end{eqnarray}
   where $T^0_{\rm R}$ is the present temperature of $\nu^{}_{\alpha {\rm R}}$ and $\overline{\nu}^{}_{\alpha {\rm L}}$. For the decoupling temperature $T^{}_{\rm R} \approx 200~{\rm MeV}$, the cosmological bound on $\Delta N^{}_{\rm eff} \equiv N^{}_{\rm eff} - 3.046 < 0.53$ is saturated and we have $T^0_{\rm R} \approx 1.28~{\rm K}$. The average number density of $\nu^{}_{\alpha {\rm R}}$ or $\overline{\nu}^{}_{\alpha {\rm L}}$ turns out to be $\overline{n}^{}_\nu(\nu^{}_{\alpha {\rm R}}) \approx 16~{\rm cm}^{-3}$, which should be compared with $\overline{n}^{}_\nu(\nu^{}_{\alpha {\rm L}}) \approx 56~{\rm cm}^{-3}$ in the standard case. Thus, in the thermal (TH) scenario, only $T^0_{\rm R} \lesssim 1.28~{\rm K}$ is allowed by cosmological observations.

\item {\it Nonthermal Right-handed Dirac Neutrinos}---In Ref.~\cite{Chen:2015dka}, it has been argued that the coupling of right-handed neutrinos to the inflaton field could give rise to a nonthermal production of $\nu^{}_{\alpha {\rm R}}$ and $\overline{\nu}^{}_{\alpha {\rm L}}$ in the reheating epoch after inflation. In this nonthermal (NT) scenario, both $\nu^{}_{\alpha {\rm R}}$ and $\overline{\nu}^{}_{\alpha {\rm L}}$ have never been in thermal equilibrium, the distribution function in the C$\nu$B frame can be written as
    \begin{eqnarray}
    f^{}_{\rm NT}(p^{}_\nu) = \left\{ \begin{array}{ccc}
                                        \eta  & \; , & p^{}_\nu \leq \varepsilon^0_{\rm F} \; ; \\
                                        ~ & ~ & ~ \\
                                        0 & \; , & p^{}_\nu > \varepsilon^0_{\rm F} \; ,
                                      \end{array}
    \right.
    \label{eq:nt}
    \end{eqnarray}
    where $\varepsilon^0_{\rm F}$ denotes the Fermi energy at present. Here $0 \leq \eta \leq 1$ is the fraction of occupied states below the Fermi energy, and $\eta = 1$ corresponds to the case of a degenerate spectrum. Without specifying any dynamics of particle production from inflaton decays, the authors of Ref.~\cite{Chen:2015dka} have assumed that the Fermi energy at production is $\varepsilon^{}_{\rm F} = \xi T^{}_{\rm R}$, where $T^{}_{\rm R}$ is the temperature of thermal bath and $\xi$ is a free parameter. The cosmological bound $\Delta N^{}_{\rm eff} < 0.53$ can be translated into a restrictive constraint on this free parameter $\xi \lesssim 3$ for a degenerate spectrum with $\eta = 1$.\footnote{To be consistent with the TH scenario, we use the same upper bound $\Delta N^{}_{\rm eff} < 0.53$ on extra neutrino species, while $\Delta N^{}_{\rm eff} < 0.7$ was adopted in Ref.~\cite{Chen:2015dka}. Therefore, a slight difference between the number density $\overline{n}^{}_\nu(\nu^{}_{\alpha {\rm R}}) \approx 29~{\rm cm}^{-3}$ in the present work and $\overline{n}^{}_\nu(\nu^{}_{\alpha {\rm R}}) \approx 36~{\rm cm}^{-3}$ in Ref.~\cite{Chen:2015dka} should be noted.} If this upper bound is saturated, today we have the Fermi energy $\varepsilon^0_{\rm F} \approx 2.75~{\rm K}$, and the average number density of $\nu^{}_{\alpha {\rm R}}$ or $\overline{\nu}^{}_{\alpha {\rm L}}$ is $\overline{n}^{}_\nu(\nu^{}_{\alpha {\rm R}}) \approx 29~{\rm cm}^{-3}$, which should be compared with $\overline{n}^{}_\nu(\nu^{}_{\alpha {\rm L}}) \approx 56~{\rm cm}^{-3}$ in the standard case. For a nondegenerate spectrum with $\eta < 1$, the number density will be slightly smaller.
\end{itemize}
Notice that the cosmological bound on $\Delta N^{}_{\rm eff} < 0.53$ will be further improved in the future. For instance, if the upper bound on $\Delta N^{}_{\rm eff}$ is reduced by a factor of two, the allowed number density of right-handed neutrinos $\overline{n}^{}_\nu(\nu^{}_{\alpha \rm R})$ will be multiplied by $2^{-3/4} \approx 0.59$ for both TH and NT cases.
\begin{figure}[t!]
\centering
\includegraphics[width=3.5in]{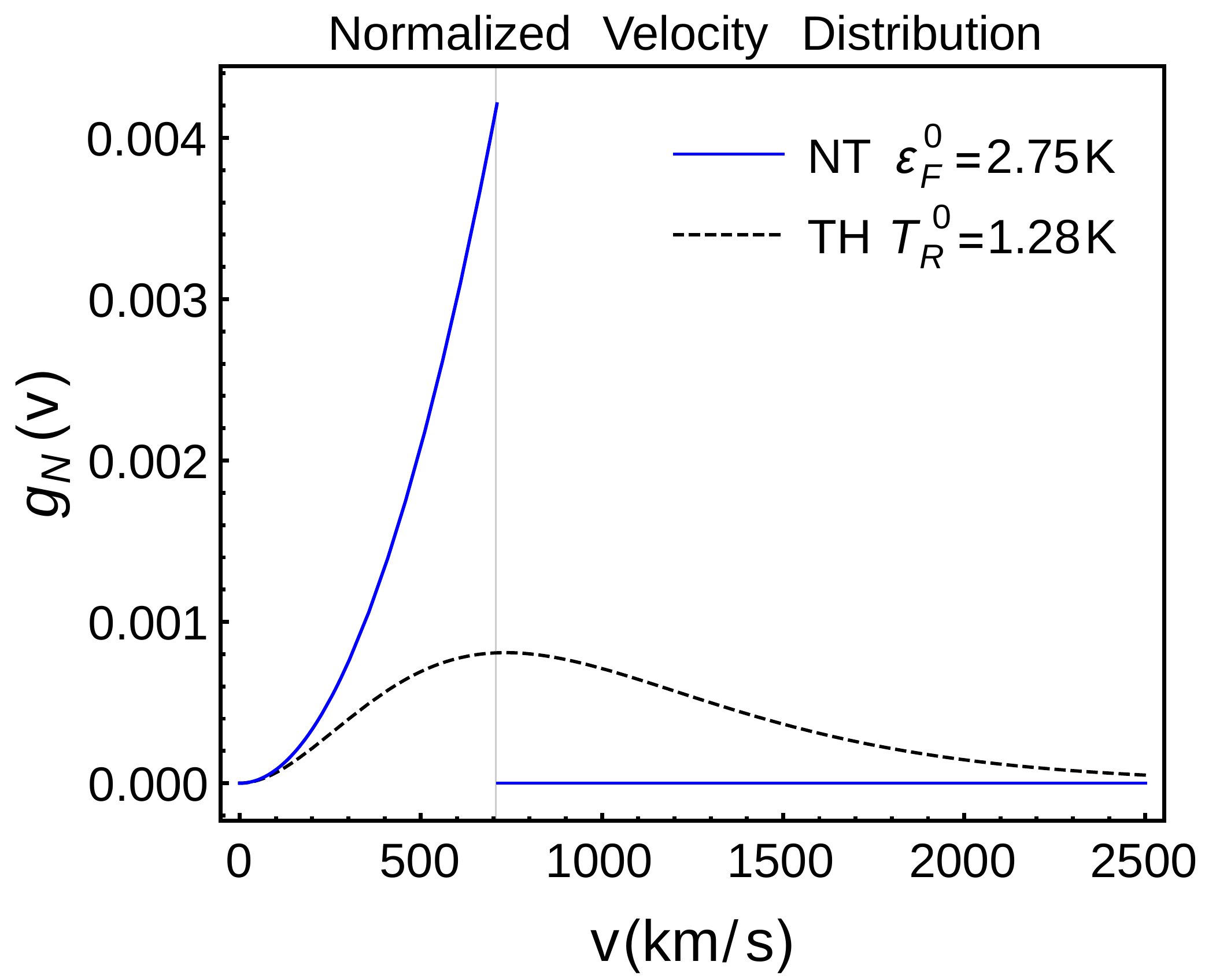}
\caption{The normalized velocity distribution $g^{}_{\rm N}(v^{}_\nu) \equiv v^2_\nu g(v^{}_\nu) m^3_\nu/(2\pi^2 \overline{n}^{}_\nu)$ for thermal and nonthermal relic right-handed Dirac neutrinos in the C$\nu$B frame, where the temperature $T^0_{\rm R} = 1.28~{\rm K}$ is taken for the former case and the Fermi energy $\varepsilon^0_{\rm F} = 2.75~{\rm K}$ for the latter. Note that the absolute neutrino mass $m^{}_\nu = 0.1~{\rm eV}$ is assumed for illustration and the speed of light $c = 3.0 \times 10^5~{\rm km}~{\rm s}^{-1}$ is used.}
\label{fig:velocity}
\end{figure}

Throughout this work, we shall take the absolute scale of neutrino masses to be $m^{}_\nu \approx 0.1~{\rm eV}$, which is marginally compatible with the cosmological bound $\sum \equiv m^{}_1 + m^{}_2 + m^{}_3 < 0.23~{\rm eV}$ on the sum of neutrino masses.\footnote{It is worthwhile to mention that a more stringent bound $\Sigma < 0.12~{\rm eV}$ or $< 0.13~{\rm eV}$ at the $95\%$ confidence level has been derived by combing the Planck CMB and Baryon Acoustic Oscillation data with the Lyman-alpha clustering data~\cite{Palanque-Delabrouille:2015pga} or with the SDSS-DR7 sample of Luminous Red Galaxies~\cite{Cuesta:2015iho}. Once such a tight bound is confirmed, the PTOLEMY experiment will have to greatly improve its energy resolution to well separate the signal from a huge background. In addition, neutrino masses below $m^{}_\nu \approx 0.1~{\rm eV}$ imply larger neutrino velocities, reducing the annual modulation, as we will discuss in the next section.} Hence, cosmic relic neutrinos are nowadays non-relativistic particles. For later convenience, we write the distribution function of non-relativistic relic neutrinos as
\begin{eqnarray}
f(p^{}_\nu) \frac{{\rm d}^3 p^{}_\nu}{(2\pi)^3} = m^3_\nu \frac{{\rm d}^3 v^{}_\nu}{(2\pi)^3} g(v^{}_\nu) \; ,
\label{eq:dist}
\end{eqnarray}
where $p^{}_\nu \approx m^{}_\nu v^{}_\nu$ holds as an excellent approximation and $g(v^{}_\nu)$ is the velocity distribution function in the C$\nu$B frame.

The integration of the right-hand side of Eq.~(\ref{eq:dist}) over neutrino velocities gives us the neutrino number density $\overline{n}^{}_\nu$. In order to compare between TH and NT scenarios, we introduce the normalized velocity distribution as $g^{}_{\rm N}(v^{}_\nu) \equiv  v^2_\nu g(v^{}_\nu) m^3_\nu/(2\pi^2 \overline{n}^{}_\nu)$ and show it for right-handed neutrinos in Fig.~1, where one can observe the main difference between these two types of spectra. The NT spectrum concentrates more on the velocity region of $200~{\rm km}~{\rm s}^{-1}$ to $700~{\rm km}~{\rm s}^{-1}$ in the C$\nu$B frame, while the TH spectrum extends to a much higher velocity.

\section{Annual Modulation}

In the standard case, it has been proposed in Ref.~\cite{Safdi:2014rza} that the gravitational focusing of C$\nu$B by the Sun will modify the distribution function of $\nu^{}_{\alpha {\rm L}}$ at the Earth and cause an annual modulation of the capture rate of C$\nu$B in the PTOLEMY experiment. We shall apply this interesting idea to the relic right-handed Dirac neutrinos $\nu^{}_{\alpha {\rm R}}$ and examine if it is possible to distinguish between TH and NT spectra, arising from different production mechanisms. In this section, we first summarize the capture rates of cosmic relic neutrinos for a PTOLEMY-like experiment. Then, following Ref.~\cite{Safdi:2014rza}, we introduce the measure of annual modulation of the capture rate for standard C$\nu$B and compare it with those for STD+TH and STD+NT with relic right-handed Dirac neutrinos.\footnote{To avoid any confusion, we emphasize that TH and NT are referring only to right-handed neutrinos, while STD+TH and STD+NT to both left- and right-handed neutrinos.}

\subsection{Capture Rates}

The capture rate of a polarized neutrino state on the tritium target was first correctly calculated in Ref.~\cite{Long:2014zva}. See Ref.~\cite{Lisanti:2014pqa} for a similar calculation, and also some discussions on measuring neutrino anisotropy on the polarized tritium target. Considering an unpolarized tritium target and a neutrino mass eigenstate $\nu^{}_i$ with a spin $s^{}_{\nu}$ being either $+1/2$ or $-1/2$, one can find the product of the cross section and the neutrino velocity $v^{}_{\nu^{}_i}$~\cite{Long:2014zva}
\begin{eqnarray}
\sigma^{}_i(s^{}_{\nu}) v^{}_{\nu^{}_i} = {\cal A}(s^{}_\nu) |U^{}_{ei}|^2  \overline{\sigma} \; ,
\label{eq:rate}
\end{eqnarray}
where $\overline{\sigma} \approx 3.8\times 10^{-45}~{\rm cm}^2$ and ${\cal A}(s^{}_\nu) \equiv 1 - 2s^{}_\nu v^{}_{\nu^{}_i}$, and $U^{}_{e i}$ for $i = 1, 2, 3$ are the three elements in the first row of the lepton flavor mixing matrix. Since ${\cal A}(+1/2) \approx 0$ and ${\cal A}(-1/2) \approx 2$ hold in the relativistic limit $v^{}_{\nu^{}_i} \to 1$, only the left-helical neutrinos can be captured. However, in the non-relativistic limit $v^{}_{\nu^{}_i} \to 0$, we have ${\cal A}(+1/2) \approx {\cal A}(-1/2) \approx 1$, so both left- and right-helical neutrino states contribute to the total capture rate. Summing over the contributions from all three neutrino mass eigenstates, we obtain the total capture rate~\cite{Long:2014zva}
\begin{eqnarray}
\Gamma^{}_{\rm D} = N^{}_{\rm T} \sum^3_{i = 1} \int \frac{{\rm d}^3 p^{}_{\nu^{}_i}}{(2\pi)^3}\left[\sigma^{}_i(-\frac{1}{2}) v^{}_{\nu^{}_i} f^{}_{\nu^{}_{\rm hL}}(p^{}_{\nu^{}_i}) + \sigma^{}_i(+\frac{1}{2}) v^{}_{\nu^{}_i} f^{}_{\nu^{}_{\rm hR}}(p^{}_{\nu^{}_i}) \right] \; ,
\label{eq:totrate}
\end{eqnarray}
where $N^{}_{\rm T}$ stands for the number of tritium nuclei, $f^{}_{\nu^{}_{\rm hL}}(p^{}_{\nu^{}_i})$ and $f^{}_{\nu^{}_{\rm hR}}(p^{}_{\nu^{}_i})$ denote the distribution functions of left-helical and right-helical neutrino states at the Earth, respectively.

In the standard scenario of massive Dirac neutrinos, the production of right-handed neutrinos $\nu^{}_{\alpha {\rm R}}$ in the early Universe is so inefficient that all the neutrino states are left-handed in chirality. In the epoch of neutrino decoupling, neutrinos are relativistic particles, so almost all of them are left-helical states. As the helicities are conserved during the expansion of our Universe, we have the number density of left-helical neutrino states $n^i_{\nu^{}_{\rm hL}} \approx 56~{\rm cm}^{-3}$ nowadays, where all the neutrino mass eigenstates are assumed to be equally populated. Meanwhile, the number density of right-helical neutrino states is vanishingly small, i.e., $n^i_{\nu^{}_{\rm hR}} \approx 0$. Following the same arguments, we arrive at $n^i_{\overline{\nu}^{}_{\rm hR}} \approx 56~{\rm cm}^{-3}$ and $n^i_{\overline{\nu}^{}_{\rm hL}} \approx 0$ for antineutrinos. Since $\sigma^{}_i v^{}_{\nu^{}_i}$ is independent of neutrino velocities in the leading-order approximation for non-relativistic neutrinos (i.e., $v^{}_{\nu^{}_i} \ll 1$), the integration over distribution functions just gives the neutrino number density. With the help of the unitarity condition of lepton flavor mixing matrix $|U^{}_{e1}|^2 + |U^{}_{e2}|^2 +|U^{}_{e3}|^2 = 1$, the capture rate of $\nu^{}_e + {^3}{\rm H} \to {^3}{\rm He} + e^-$ at PTOLEMY is found to be $\Gamma^{}_{\rm D} \approx 4~{\rm yr}^{-1}$~\cite{Long:2014zva}.

In the nonstandard scenario of either TH or NT production mechanism, the right-handed neutrinos $\nu^{}_{\alpha {\rm R}}$ and left-handed antineutrinos $\overline{\nu}^{}_{\alpha {\rm L}}$ can be populated, although their number densities are well constrained by cosmological observations. In assumption of equal number densities for neutrino mass eigenstates, we obtain at most $n^i_{\nu^{}_{\rm hR}} = n^i_{\overline{\nu}^{}_{\rm hL}} \approx 16~{\rm cm}^{-3}$ and $n^i_{\nu^{}_{\rm hR}} = n^i_{\overline{\nu}^{}_{\rm hL}} \approx 29~{\rm cm}^{-3}$ for TH and NT cases, respectively, depending on the decoupling temperature $T^{}_{\rm R}$ in the former case while the initial Fermi energy $\varepsilon^{}_{\rm F}$ and the occupation fraction $\eta$ in the latter. Neglecting the velocity dependence and inserting these number densities into Eq.~(\ref{eq:totrate}), one can get the total capture rate $\Gamma^{}_{\rm D} \approx 5.1~{\rm yr}^{-1}$ and $\Gamma^{}_{\rm D} \approx 6.1~{\rm yr}^{-1}$ for STD+TH and STD+NT scenarios, as already mentioned before.

\subsection{Focusing Effects}

To discriminate the velocity distributions of relic neutrinos, as shown in Fig.~\ref{fig:velocity}, now we should take into account all possible dependence on neutrino velocities when calculating the capture rates. First, the spin factor ${\cal A}(s^{}_\nu)$ in Eq.~(\ref{eq:rate}) actually depends on neutrino velocities, namely, ${\cal A}(s^{}_\nu) = 1 + v^{}_\nu$ for left-helical states and ${\cal A}(s^{}_\nu) = 1 - v^{}_\nu$ for right-helical states.\footnote{If massive neutrinos are Majorana particles, the distribution functions of left-helical and right-helical neutrino states are identical and thus the velocity dependence in ${\cal A}(s^{}_\nu)$ is completely cancelled out~\cite{Safdi:2014rza}.} Although such corrections to the total rate are insignificant, we include them for completeness.

Second, the impact of the Sun's gravity on the dark matter distribution has already been investigated in Refs.~\cite{Griest:1987vc,Alenazi:2006wu,Lee:2013xxa}. As has been pointed out in Ref.~\cite{Safdi:2014rza}, even if the relic neutrinos are unbound to the Milky Way, the gravity of the Sun will modify the velocity distribution of C$\nu$B at the Earth, where the detectors are located. In connection with gravitational effects on neutrinos, it is worthwhile to notice that massive neutrinos are subject to gravitational clustering in the dark-matter halo of the galactic scale, which has been studied carefully by following the nonlinear evolution of C$\nu$B in Refs.~\cite{Ringwald:2004np,VillaescusaNavarro:2012ag}.
For $m^{}_\nu = 0.1~{\rm eV}$ the relic neutrino overdensity $n^{}_\nu/\overline{n}^{}_\nu - 1$ is found to be more than $10\%$ in Ref.~\cite{VillaescusaNavarro:2012ag}, while for $m^{}_\nu = 0.15~{\rm eV}$ a value about $40\%$ is reached in Ref.~\cite{Ringwald:2004np}. Thus, it may be reasonable to expect an enhancement of $30\%$ in the local neutrino density. On the other hand, some deviations from the Fermi-Dirac distribution have also been observed~\cite{Ringwald:2004np,VillaescusaNavarro:2012ag}. As shown in Ref.~\cite{VillaescusaNavarro:2012ag}, the proportion of neutrinos with peculiar velocities below $100~{\rm km}~{\rm s}^{-1}$ is smaller by a factor of $1.33$ than that for the Fermi-Dirac distribution. The impact on higher velocities will be much smaller. Though we ignore these effects in our calculations, a relative error about $30\%$ should be kept in mind when we calculate the capture rate and annual modulation.
\begin{figure}[t!]
\centering
\hspace{0.5cm}
\includegraphics[width=3.9in]{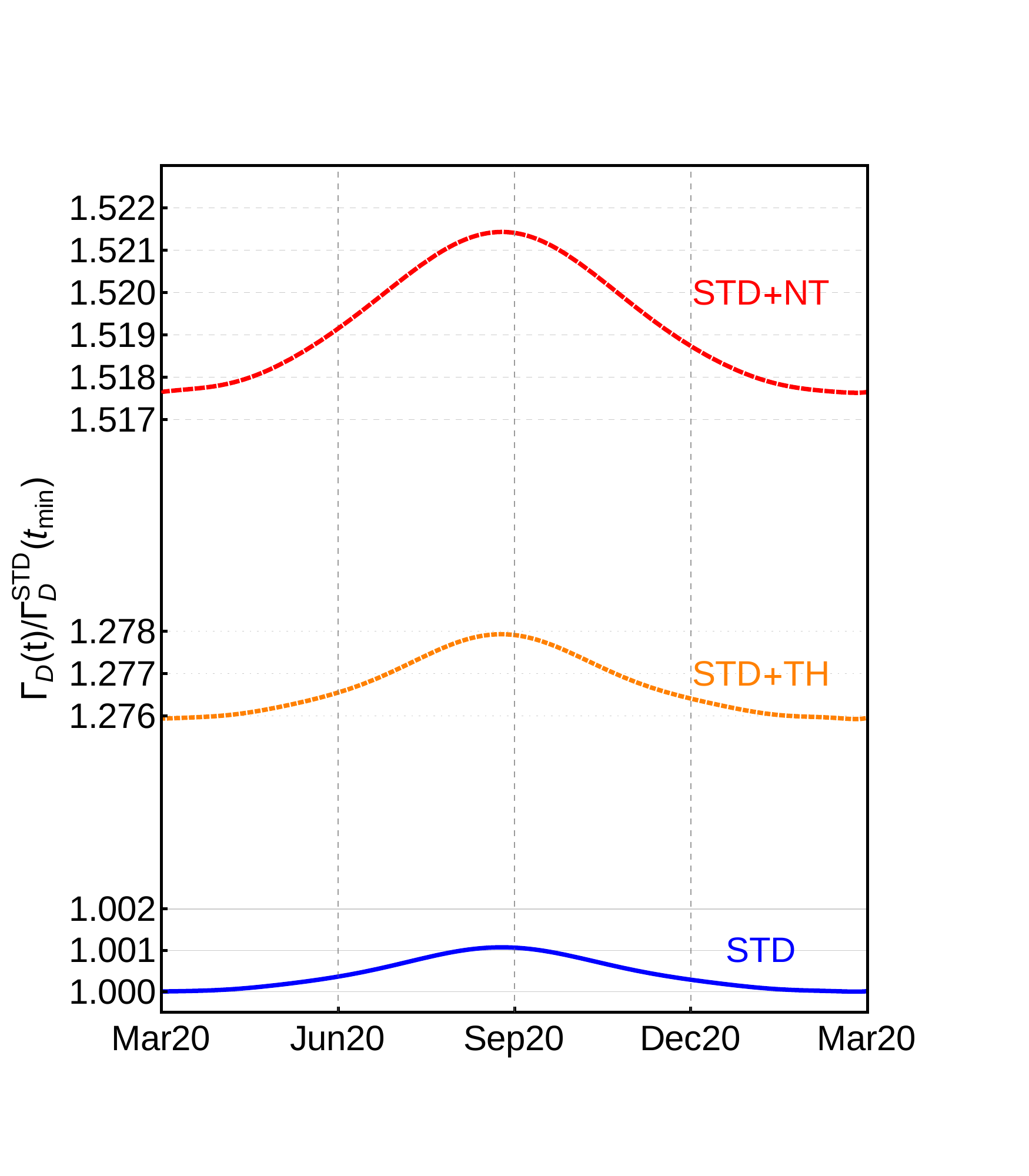}
\vspace{-0.8cm}
\caption{The normalized capture rate $\Gamma^{}_{\rm D}(t)/\Gamma^{\rm STD}_{\rm D}(t^{}_{\rm min})$ of relic neutrinos in the PTOLEMY experiment has been shown for the standard (STD) scenario (blue solid curve), the nonstandard scenario of thermal (TH) right-handed Dirac neutrinos (orange dotted curve) with $T^0_{\rm R} = 1.28~{\rm K}$, and the case of nonthermal (NT) right-handed Dirac neutrinos (red dashed curve) with $\varepsilon^0_{\rm F} = 2.75~{\rm K}$., where $\Gamma^{\rm STD}_{\rm D}(t^{}_{\rm min})$ denotes the capture rate at $t^{}_{\rm min}$ in the standard case.}
\label{fig:totrate}
\end{figure}

In order to derive the distribution function of relic neutrinos relative to the Earth, which will be denoted as $g^{}_{\oplus}({\bf v}^{}_\nu)$, we have to know the velocity of the Earth ${\bf V}^{}_\oplus$ relative to the Sun. Working in the heliocentric ecliptic coordinate system, for which the ecliptic plane is spanned by two unit vectors $\hat{\bf e}{}_1$ and $\hat{\bf e}^{}_2$, one can find ${\bf V}^{}_\oplus = V^{}_\oplus \left[\hat{\bf e}^{}_1 \cos\omega(t - t^{}_{\rm ve}) + \hat{\bf e}^{}_2 \sin\omega(t-t^{}_{\rm ve})\right]$ with $V^{}_\oplus = 29.79~{\rm km}~{\rm s}^{-1}$ and $\omega = 2\pi~{\rm yr}^{-1}$, where $t^{}_{\rm ve} \approx {\rm March}~20$ is the time of the vernal equinox~\cite{Lee:2013xxa,McCabe:2013kea}. The velocity of the Sun relative to the C$\nu$B frame is approximately given by ${\bf V}^{}_\odot = V^{}_\odot (-0.0695, -0.662, 0.747)$ with $V^{}_\odot = 369~{\rm km}~{\rm s}^{-1}$. The velocity ${\bf V}^{}_\odot$ can be identified from the dipole structure of CMB~\cite{Aghanim:2013suk} if we approximately identify the CMB frame with C$\nu$B frame, which is quite reasonable as no significant clustering of neutrinos takes place in the epoch of CMB formation~\cite{Safdi:2014rza}. Finally, due to the Sun's gravity, the neutrino velocity ${\bf v}^{}_\nu$ at the Earth corresponds to the velocity ${\bf v}^{}_\infty$ at an infinitely far distance to the Sun
\begin{eqnarray}
{\bf v}^{}_\infty ({\bf v}^{}_\nu) = \frac{v^2_\infty {\bf v}^{}_\nu + v^{}_\infty (G^{}_{\rm N} M^{}_{\odot}/|{\bf r}|)\hat{\bf r} - v^{}_\infty ({\bf v}^{}_\nu \cdot \hat{\bf r}) {\bf v}^{}_\nu }{ v^2_\infty + G^{}_{\rm N} M^{}_{\odot}/|{\bf r}| - v^{}_\infty ({\bf v}^{}_\nu\cdot \hat{\bf r}) } \; ,
\label{eq:vinfty}
\end{eqnarray}
where $v^{}_\infty \equiv |{\bf v}^{}_\infty|$ and $\hat{\bf r} \equiv {\bf r}/|{\bf r}|$ with ${\bf r}$ being the displacement vector of the Earth, $G^{}_{\rm N}$ Newton's gravitational constant, and $M^{}_\odot$ the solar mass. According to the Liouville theorem, the distribution function is constant along any trajectory in the phase space. Thus, we have
\begin{eqnarray}
g^{}_\oplus({\bf v}^{}_\nu) = g\left[{\bf V}^{}_\odot + {\bf v}^{}_\infty({\bf V}^{}_\oplus + {\bf v}^{}_\nu)\right] \; ,
\label{eq:distE}
\end{eqnarray}
for the local distribution function of relic neutrinos at the Earth. As is shown in Ref.~\cite{Safdi:2014rza}, the minimal neutrino density in the standard case is reached around $t^{}_{\rm min} = t^{}_{\rm ev} - 8~{\rm days}$ (namely, March 12) when the Earth is most upwind, while the maximum is achieved half a year later, i.e., on September 11.

Now we can insert the distribution function $g^{}_\oplus({\bf v}^{}_\nu)$ for relic neutrinos in Eq.~(\ref{eq:distE}) together with Eq.~(\ref{eq:dist}) into the total capture rate in Eq.~(\ref{eq:totrate}), and derive the rates at an arbitrary time during the year both in the standard scenario and in two nonstandard scenarios with right-handed neutrinos. The final results of $\Gamma^{}_{\rm D}(t)/\Gamma^{\rm STD}_{\rm D}(t^{}_{\rm min})$ are illustrated in Fig.~\ref{fig:totrate}, where all the total capture rates $\Gamma^{}_{\rm D}(t)$ have been normalized to the minimal rate $\Gamma^{\rm STD}_{\rm D}(t^{}_{\rm min})$ on March 12 in the standard case. Since $\sigma^{}_i v^{}_{\nu^{}_i}$ is almost velocity-independent, the normalized capture rates directly reflect the local neutrino number densities along the orbit of the Earth. 

Two main differences between standard and nonstandard cases can be observed. First, the relic right-handed neutrinos in STD+TH and STD+NT scenarios contribute to the capture rate by a sizable amount of $28\%$ and $52\%$, respectively. Second, the difference between maximum and minimum increases remarkably in the two nonstandard cases, where right-handed neutrinos concentrate more on the region of low velocities and thus they are affected greatly by the Sun's gravity. To be explicit, the average neutrino velocity in the STD case is approximately $1589~{\rm km}~{\rm s}^{-1}$ for $m^{}_\nu = 0.1~{\rm eV}$, while the average velocities of right-handed neutrinos for TH and NT cases are around $1043~{\rm km}~{\rm s}^{-1}$ and $533~{\rm km}~{\rm s}^{-1}$, respectively. So the right-handed neutrinos in the STD+TH and STD+NT scenarios will increase the fraction of low-velocity neutrinos, when compared to the STD case.

\subsection{Detailed Comparison}

\begin{figure}[!t]
\begin{center}
\subfigure{%
\includegraphics[width=0.455\textwidth]{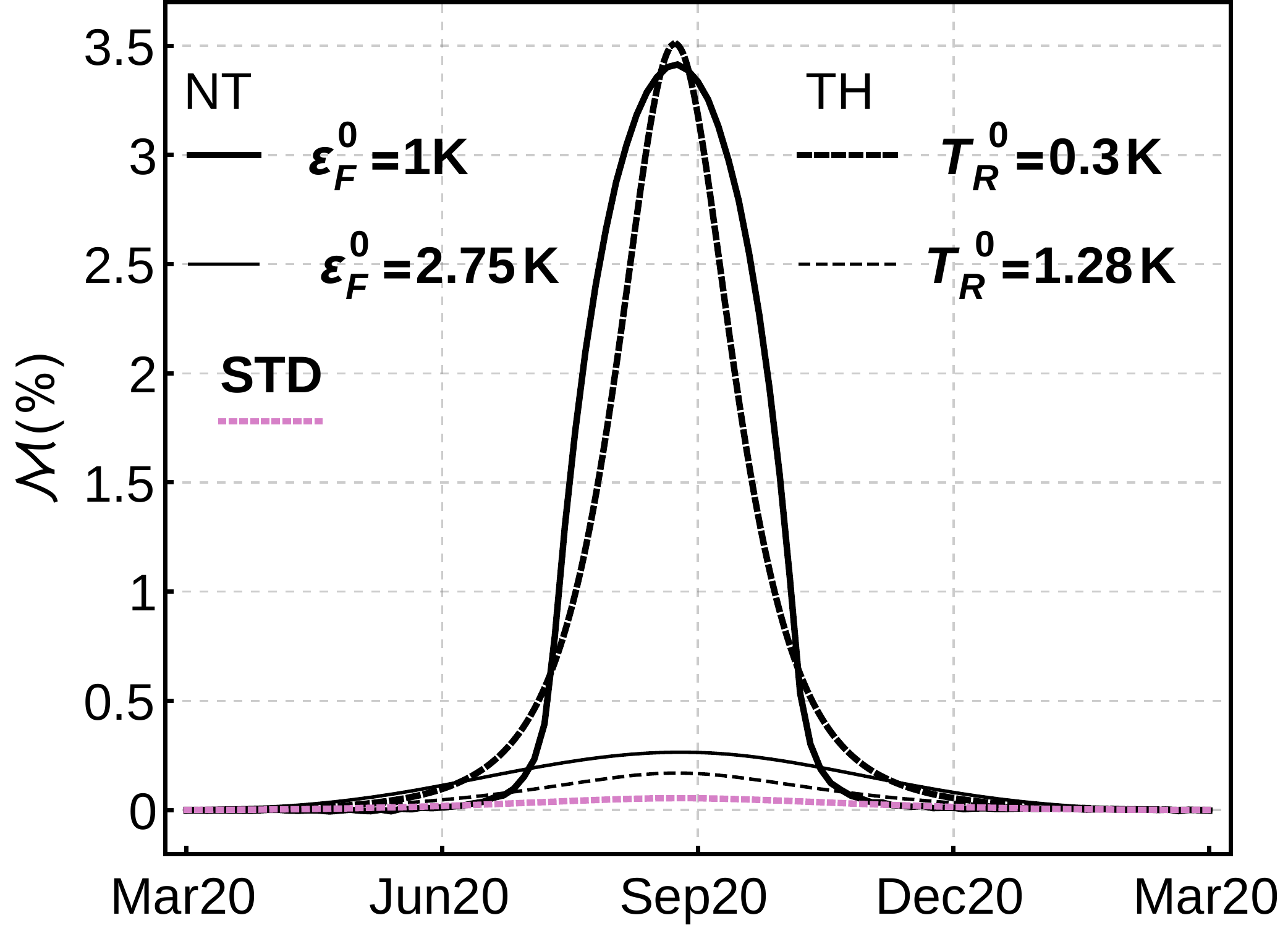}        }%
\subfigure{%
\hspace{0.8cm}
\includegraphics[width=0.464\textwidth]{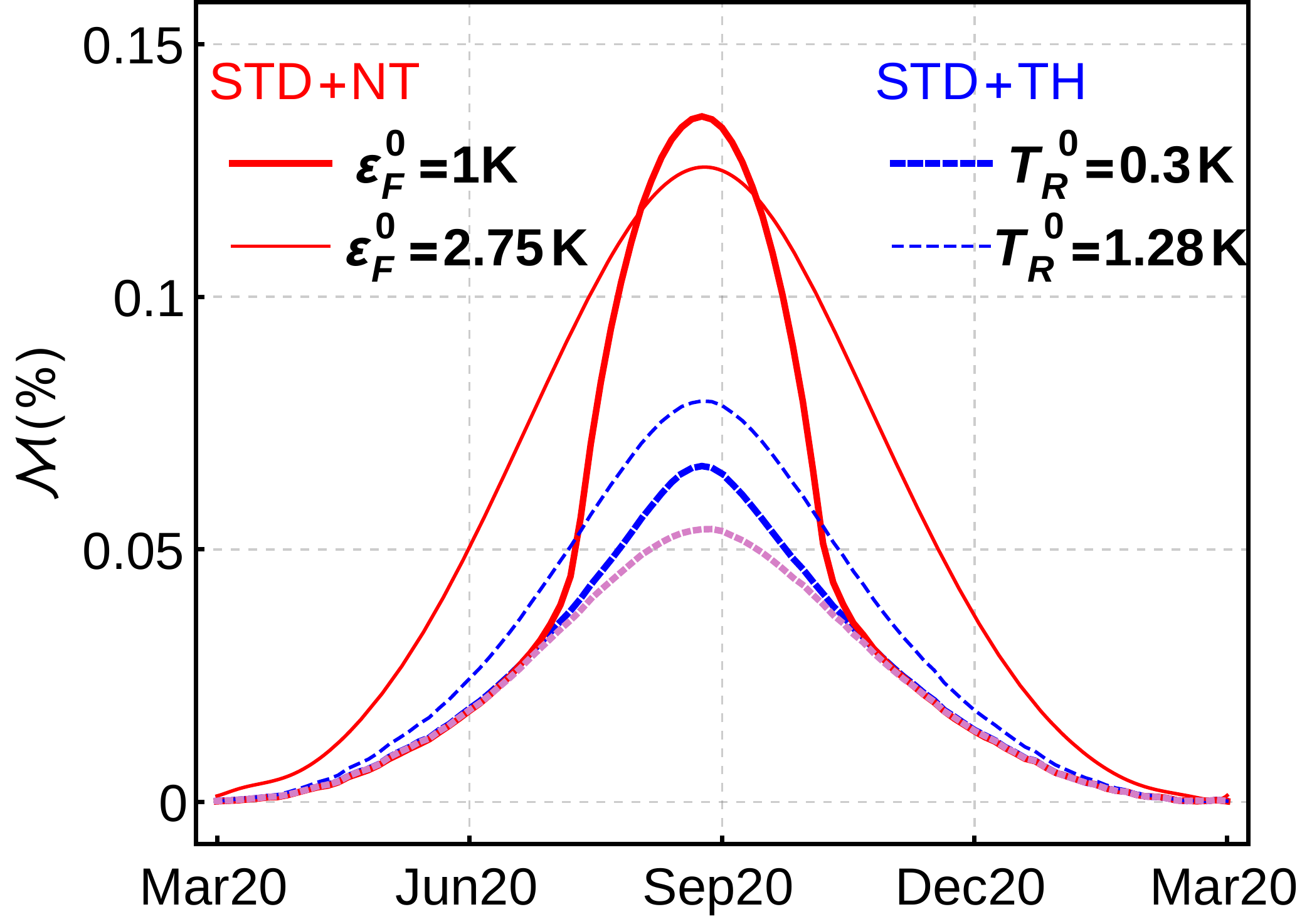}        }
\end{center}
\vspace{-0.3cm}
\caption{The annual modulation ${\cal M}(t)$ of the C$\nu$B capture rate in the PTOLEMY experiment, where the modulation for the standard case (STD) is represented by the magenta dotted curve, while those for TH and NT cases by the dashed and solid curves, respectively. On the left panel, the modulation is shown only for right-handed neutrinos, and two different values $T^0_{\rm R} = 0.3~{\rm K}$ and $T^0_{\rm R} = 1.28~{\rm K}$ are chosen for the TH case while $\varepsilon^0_{\rm F} = 1~{\rm K}$ and $\varepsilon^0_{\rm F} = 2.75~{\rm K}$ for the NT case. On the right panel, the contributions from both left-handed and right-handed neutrinos to the capture rates are combined.}
\label{fig:comparison}
\end{figure}

To make a detailed comparison, we follow Ref.~\cite{Safdi:2014rza} and define the annual modulation as
\begin{eqnarray}
{\cal M}(t) \equiv \frac{\Gamma^{}_{\rm D}(t) - \Gamma^{}_{\rm D}(t^{}_{\rm min})}{\Gamma^{}_{\rm D}(t) + \Gamma^{}_{\rm D}(t^{}_{\rm min})} \; ,
\label{eq:modulation}
\end{eqnarray}
where $t^{}_{\rm min} \approx {\rm March}~12$ stands for the time of minimal capture rate. In Fig.~\ref{fig:comparison}, the annual modulation has been calculated for both standard and nonstandard scenarios. On the left panel, the modulation for the standard C$\nu$B is plotted as a magenta dotted curve, from which one can read the maximum is only about $0.05\%$ (See also the same curve zoomed in on the right panel). In the TH case, the thick and thin dashed curves correspond to the modulations for the temperature of $T^0_{\rm R} = 0.3~{\rm K}$ and $T^0_{\rm R} = 1.28~{\rm K}$, respectively. Similar results for the NT case of $\varepsilon^0_{\rm F} = 1~{\rm K}$ and $\varepsilon^0_{\rm F} = 2.75~{\rm K}$ are denoted by thick and thin solid curves. On the right panel, the combined contributions from both left-handed and right-handed neutrinos to the modulation have been
depicted. Some comments are in order:
\begin{itemize}
\item One can observe from the left panel of Fig.~\ref{fig:comparison} that the peak modulation can be as large as $3.5\%$ in the TH case for $T^0_{\rm R} = 0.3~{\rm K}$ and in the NT case for $\varepsilon^0_{\rm F} = 1~{\rm K}$, which should be compared to ${\cal M} \lesssim 0.05\%$ in the standard case. For the other model parameters $T^0_{\rm R} = 1.28~{\rm K}$ and $\varepsilon^0_{\rm F} = 2.75~{\rm K}$, for which the number density of right-handed neutrinos saturates the cosmological upper bound on $\Delta N^{}_{\rm eff}$, the maximal modulations in both TH and NT will be reduced by a factor of twenty. This observation demonstrates that the impact of the Sun's gravity on the local density of relic neutrinos is more important for slower neutrinos.

\item In reality, the modulation should receive the contributions from both left-handed neutrinos (i.e., the standard C$\nu$B) and right-handed neutrinos, and one cannot turn off the former contribution. Therefore, only the curves labelled by STD+TH and STD+NT on the right panel correspond to realistic situations. On the left panel, the modulations solely from right-handed neutrinos are given to illustrate their main differences from the result in the standard case. Although the amplitude of annual modulation due to right-handed neutrinos is larger for a lower temperature or Fermi energy, their number density will be smaller, leading to a reduced capture rate. This is why the peak modulations for different model parameters (namely, $T^0_{\rm R}$ for TH and $\varepsilon^0_{\rm F}$ for NT) on the right panel are now comparable.
\end{itemize}
Finally, we consider if it is possible to distinguish between STD+TH and STD+NT scenarios, assuming the number density of right-handed neutrinos to be the same. This consideration does make sense, since the total capture rates will be distinguishable if the number densities of right-handed neutrinos are different. Hence, in Fig.~\ref{fig:density}, we show the maximal modulation for a given number density $N^{}_{\rm R}$ of right-handed neutrinos, which has been normalized to the present-day number density of CMB photons $N^{}_\gamma \approx 411~{\rm cm}^{-3}$. In either TH or NT scenario, once a value of $T^0_{\rm R}$ or $\varepsilon^0_{\rm F}$ is assumed, one can immediately calculate the corresponding number density $N^{}_{\rm R}$ and the capture rate $\Gamma^{}_{\rm R}$, which are actually displayed on the bottom and top axes. Some important observations related to Fig.~\ref{fig:density} should be emphasized:
\begin{figure}[!t]
\begin{center}
\subfigure{%
\includegraphics[width=0.415\textwidth]{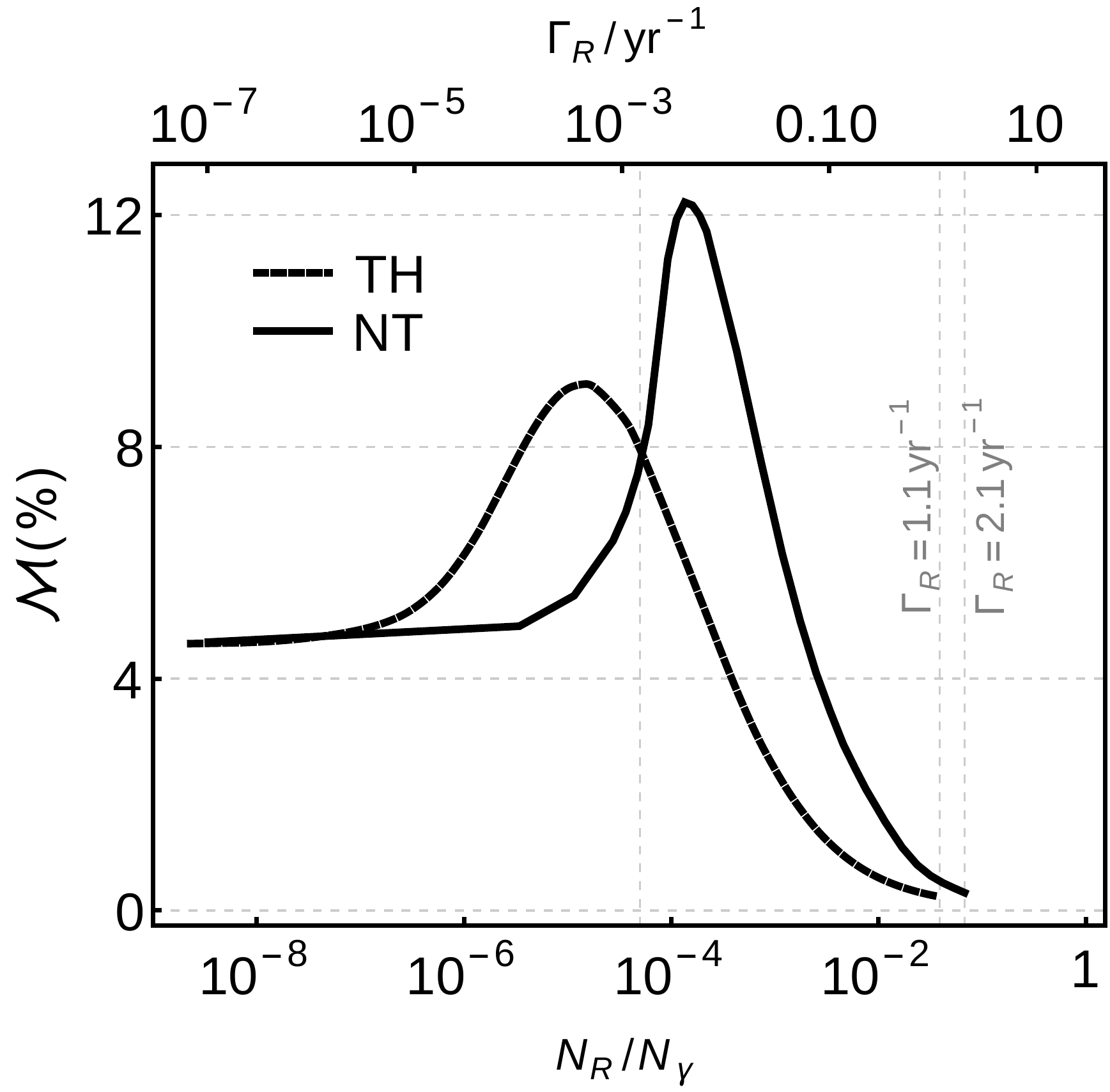}        }%
\subfigure{%
\hspace{0.8cm}
\includegraphics[width=0.438\textwidth]{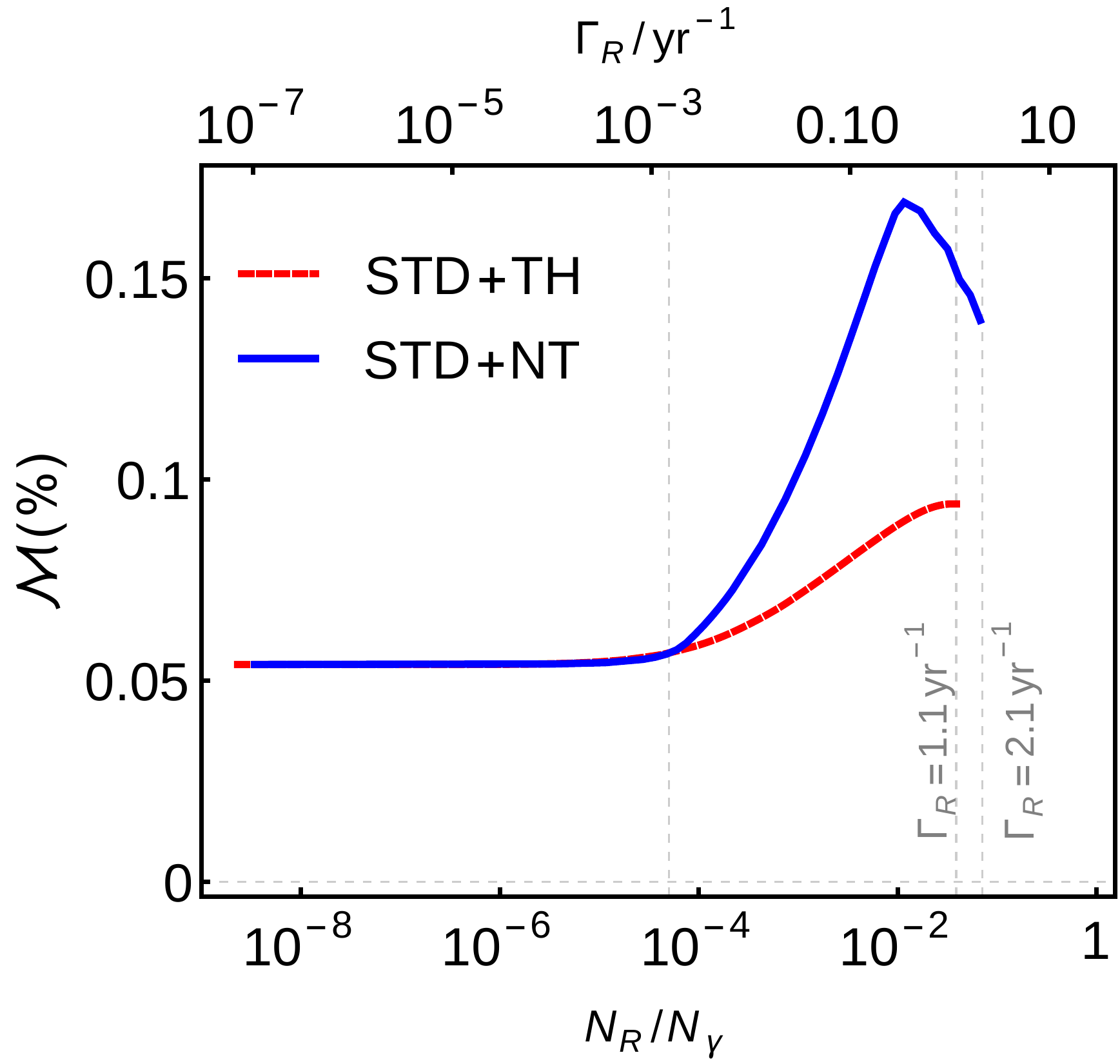}        }
\end{center}
\vspace{-0.5cm}
\caption{The amplitude of the modulation ${\cal M}$ is shown with respect to the ratio of right-handed neutrino number density $N^{}_{\rm R}$ to the photon number density $N^{}_\gamma$. The results for only right-handed neutrinos in two nonstandard scenarios are given on the left panel, while the combined contributions from both left-handed and right-handed neutrinos are shown on the right panel where the corresponding capture rate solely from right-handed neutrinos is given on the top axis.}
\label{fig:density}
\end{figure}
\begin{enumerate}
\item For an extremely small number density $N^{}_{\rm R}$, the contribution of right-handed neutrinos to the total capture rate is negligible, as is shown on the right panel of Fig.~\ref{fig:density}. Therefore, the modulation for STD+NT (blue solid curve) and that for STD+TH (red dashed curve) are approaching ${\cal M} \approx 0.05\%$ from right, which is the limiting value in the standard case. To the right of the vertical dashed line around $N^{}_{\rm R}/N^{}_\gamma \approx 10^{-4}$, it is evident that the modulations for STD+TH and STD+NT scenarios are clearly distinguishable.

\item It should be noted that the slowly-moving neutrino wind does not imply a large modulation for extremely low neutrino velocities. The main reason can be understood as follows. In the heliocentric ecliptic coordinate system, the wind of C$\nu$B is moving towards the solar system in an inclined angle about $0.06\pi$ or $10.8^\circ$, which indicates that the focus point through the solar gravitational lens does not exactly match the maximum point of modulation. Because of the dispersion of neutrino velocities, the focus of neutrino wind is broadened to cover the maximum point of modulation. It is straightforward to verify that this can be achieved if the neutrino velocity increases to $V^{}_\odot \times 0.06\pi \approx 70~{\rm km}~{\rm s}^{-1}$. Therefore, if neutrino velocities are too low, the focus region is far from the orbit of the Earth, implying a small modulation.

\item For TH and NT right-handed neutrinos, their number densities $N^{}_{\rm R}$ can be calculated by using the distribution functions in Eq.~(\ref{eq:th}) and Eq.~(\ref{eq:nt}), respectively. Thus, the average temperature $T^0_{\rm R}$ or Fermi energy $\varepsilon^0_{\rm F}$ of right-handed neutrinos in the TH or NT case is proportional to $N^{1/3}_{\rm R}$, implying that a larger value of $N^{}_{\rm R}$ corresponds to more faster neutrinos. On the left panel of Fig.~\ref{fig:density}, one may notice that there exists a maximum of the modulation for varying number densities $N^{}_{\rm R}$ in either NT or TH scenario. The reason is the same as above. The average velocity at the maximum point is $67~{\rm km}~{\rm s}^{-1}$ for the NT case, while $73~{\rm km}~{\rm s}^{-1}$ for the TH case, which is very close to $70~{\rm km}~{\rm s}^{-1}$ that we have found before. After these maximum points, the modulation decreases due to the fact that the neutrino wind is less deflected for increasing velocities. The two vertical lines on the right part arise from the cosmological bound on $\Delta N^{}_{\rm eff}$, constraining the number density of right-handed neutrinos from above. On the right panel, however, those maxima are no longer visible, which can be understood by the relatively small number density of relic right-handed neutrinos at the maximum compared to that of left-handed neutrinos. Note that new maxima indeed arise from the superposition of the contributions from both left-handed and right-handed neutrinos and are located at different number densities.
\end{enumerate}

It is now clear that the maximal modulation in the nonstandard STD+TH and STD+NT scenarios ranges from $0.1\%$ to $0.15\%$, which is twice or three times larger than ${\cal M} \approx 0.05\%$ in the standard case. In order to observe such an effect, we need about $10^6$ events, which can be accumulated in a PTOLEMY-like experiment with a tritium target of two ton running for ten years. In accumulating the data, we should also notice that the incoming direction of the neutrino wind is changing, leading to a drift of the maximum point of modulation at about $0.3^{\circ}$ per 20 years~\cite{McCabe:2013kea}.

\section{Concluding Remarks}

The recently proposed PTOLEMY experiment will be able to detect relic neutrinos from the Big Bang via $\nu^{}_e + {^3}{\rm H} \to {^3}{\rm He} + e^-$ by using a 100 gram of surface-deposition tritium source and an excellent energy resolution of $0.15~{\rm eV}$~\cite{Betts:2013uya}. Assuming that massive neutrinos are Dirac particles, the authors of Ref.~\cite{Zhang:2015wua} and Ref.~\cite{Chen:2015dka} have found that the capture rate of cosmic relic neutrinos can be enhanced from $\Gamma^{}_{\rm D} \approx 4~{\rm yr}^{-1}$ in the standard case to be $\Gamma^{}_{\rm D} \approx 5.1~{\rm yr}^{-1}$ and $\Gamma^{}_{\rm D} \approx 6.1~{\rm yr}^{-1}$, if right-handed neutrinos can be produced thermally and nonthermally in the early Universe. In this short note, following the idea of annual modulation induced by the Sun's gravity in Ref.~\cite{Safdi:2014rza}, we have pointed out that it is possible to distinguish between thermal and nonthermal spectra of cosmic relic neutrinos.

In the optimistic case, where the cosmological bound on the extra neutrino species $\Delta N^{}_{\rm eff} < 0.53$ is saturated, the modulation of the C$\nu$B capture rate reaches its maximum ${\cal M} \approx 0.15\%$ or ${\cal M} \approx 0.1\%$ on September 11 for the STD+NT and STD+TH scenario, respectively. This is to be compared with the maximal modulation ${\cal M} \approx 0.05\%$ in the standard case for the absolute neutrino mass scale $m^{}_\nu = 0.1~{\rm eV}$. The choice of this absolute mass is marginally compatible with the cosmological bound on the sum of three neutrino masses $\Sigma < 0.23~{\rm eV}$. For a larger neutrino mass, the amplitude of modulation will be increased and the gravitational clustering effects will enhance the local number density of cosmic relic neutrinos. In this case, however, neutrinos will be a mixture of unbound and bound components, implying an even smaller difference between STD+NT and STD+TH spectra. As shown in Ref.~\cite{Safdi:2014rza}, the bound neutrino wind has a different focal point around the Sun, implying a different phase of annual modulation. The velocity distribution of either NT or TH right-handed neutrinos will be affected significantly by gravitational clustering in the dark-matter halo of our galaxy and the difference between them will be diminished. Therefore, a relatively large neutrino mass is not always helpful.

As a rough estimate, we have to accumulate more than $10^6$ events in a PTOLEMY-like experiment to see an annual modulation of $0.1\%$ at the $2\sigma$ level~\cite{Safdi:2014rza}. Such a detection might not be impossible in the future when the PTOLEMY experiment is successful in discovering cosmic relic neutrinos and then scaled up to a $10~{\rm kg}$ target mass or more. Furthermore, the general idea presented in this note is to discriminate thermal and nonthermal energy spectra of cosmic relic particles via the annual modulation of detection rates. It can be applied to
\begin{itemize}
\item {\it eV-mass sterile neutrinos} -- They are able to explain a possible extra radiation during the BBN and CMB era, and can be implemented to relax the tension among cosmological parameters determined from data sets~\cite{Abazajian:2012ys}. Such light sterile neutrinos can be produced in the early Universe via a mixing with active neutrinos, or exotic interactions with dark matter.

\item {\it keV-mass sterile neutrinos} -- They are one of the most promising candidates for warm dark matter~\cite{Adhikari:2016bei}. In the early Universe, there are several different ways to produce keV-mass sterile neutrinos, including a mixing with active neutrinos in the presence of a large lepton number asymmetry and decays from a singlet scalar.
\end{itemize}
For these sterile neutrinos, a small mixing with active neutrinos guarantees a similar capture process and an annual modulation at the PTOLEMY experiment. However, the masses of sterile neutrinos are much larger than those of ordinary neutrinos, so it may be important to investigate their evolution in the gravitational potential and figure out their velocity distributions in the present solar system. As both the production mechanism and thermal history of such particles can be very different from ordinary neutrinos, we may even have a distinct phase in the annual modulation. We leave these interesting topics for a future work.

\section*{Acknowledgements}

This work was supported in part by the National Recruitment Program for Young Professionals and by the CAS Center for Excellence in Particle Physics (CCEPP).

\end{document}